\pdfoutput=1
\documentclass[letterpaper,aps,prl,twocolumn,groupedaddress,floatfix]{revtex4}

\usepackage{graphicx} % include figure files
\usepackage{mathptmx} % make Times Roman font for text AND math
\usepackage{amssymb} % need this for \lesssim symbol, others?
\usepackage{textcomp} % provides \textdegree symbol

\begin{document}

\title{Single-qubit-gate error below 10$^{-4}$ in a trapped ion}

\author{K. R. Brown}
\altaffiliation{kbrown@iontrapping.com; present address: Georgia Tech Research Institute, 400 10th Street Northwest, Atlanta, Georgia 30318, USA.}
\author{A. C. Wilson}
\author{Y. Colombe}
\author{C. Ospelkaus}
\altaffiliation{Present address: QUEST, Leibniz Universit\"{a}t Hannover, Im Welfengarten 1, D-30167 Hannover and PTB, Bundesallee 100, D-38116 Braunschweig, Germany.}
\author{A. M. Meier}
\author{E. Knill}
\author{D. Leibfried}
\author{D. J. Wineland}
\affiliation{National Institute of Standards and Technology, 325 Broadway, Boulder, CO 80305, USA}

\begin{abstract}
With a $^9$Be$^+$ trapped-ion hyperfine-states qubit, we demonstrate
an error probability per randomized single-qubit gate of
2.0(2)~$\times$~10$^{-5}$, below the threshold estimate of 10$^{-4}$
commonly considered sufficient for fault-tolerant quantum
computing. The $^9$Be$^+$ ion is trapped above a microfabricated
surface-electrode ion trap and is manipulated with microwaves applied to
a trap electrode. The achievement of low single-qubit-gate errors is an
essential step toward the construction of a scalable quantum computer.
\end{abstract}

\maketitle

In theory, quantum computers can solve certain problems much more
efficiently than classical computers \cite{nielsen01}. This has
motivated experimental efforts to construct and to verify devices that
manipulate quantum bits (qubits) in a variety of physical systems
\cite{ladd10}. The power of quantum computers depends on the ability
to accurately control sensitive superposition amplitudes by means of
quantum gates, and errors in these gates are a chief obstacle to
building quantum computers \cite{divincenzo00}. Small gate errors
would enable fault-tolerant operation through the use of quantum error
correction protocols \cite{preskill98}. While the maximum tolerable
error varies between correction strategies, there is a consensus that
$10^{-4}$ is an important threshold to breach
\cite{preskill98,knill10}. Single-qubit gates with errors slightly
above this level have been achieved with nuclear spins in liquid-state
nuclear-magnetic resonance experiments \cite{ryan09} and with neutral
atoms confined in optical lattices \cite{olmschenk10}; here we
demonstrate single-qubit error probabilities of $2.0(2)\times10^{-5}$,
substantially below the threshold. Reaching fault-tolerance still
requires reducing two-qubit-gate errors from the current state of the
art (7~$\times$~10$^{-3}$ for laser-based \cite{benhelm08} and 0.24
for microwave-based gates \cite{ospelkaus11}) to similar levels.

To determine the average error per gate (EPG), we use the method of
randomized benchmarking \cite{knill08}. Compared to other methods for
evaluating gate performance, such as quantum process tomography
\cite{poyatos97}, randomized benchmarking offers the advantage that it
efficiently and separately can determine the EPG and the combined
state-preparation and measurement errors. Because it involves long
sequences of random gates, it is sensitive to errors occurring when
gates are used in arbitrary computations. In randomized benchmarking,
the qubit, initialized close to a pure quantum state, is subjected to
predetermined sequences of randomly selected Clifford gates
\cite{bravyi05} for which, in the absence of errors, the measurement
outcome is deterministic and efficiently predictable. Clifford gates
include the basic unitary gates of most proposed fault-tolerant
quantum computing architectures. Together with certain single-qubit
states and measurements, they suffice for universal quantum computing
\cite{bravyi05,knill05}. To establish the EPG, the actual measurement
and predicted outcome are compared for many random sequences of
different lengths. Under assumptions presented in Ref.~\cite{knill08},
this yields an average fidelity as a function of the number of gates that
decreases exponentially to $1/2$ and determines the EPG. Randomized
benchmarking has been used to quantify single-qubit EPGs in a variety
of systems as summarized in Table~\ref{table:previous_benchmarking}.
\begin{table*}
\begin{center}
%\begin{tabular}{ | l | l | c | }
\begin{tabular*}{\textwidth}{@{\extracolsep{\fill}} l  l  r  }
\hline
\hline
Reference & System & Gate error \\
\hline
This Rapid Communication (2011) & Single trapped ion & $2.0(2)\times10^{-5}$\\
Reference \cite{ryan09} (2009) & Nuclear magnetic resonance & $1.3(1)\times10^{-4}$ \\
Reference \cite{olmschenk10} (2010) & Atoms in an optical lattice & $1.4(1)\times10^{-4}$ \\
Reference \cite{biercuk09} (2009) & Trapped-ion crystal & $8(1)\times10^{-4}$ \\
Reference \cite{knill08} (2008) & Single trapped ion & $4.8(2)\times10^{-3}$ \\
Reference \cite{chow10} (2010) & Superconducting transmon & $7(5)\times10^{-3}$ \\
\hline
\hline
\end{tabular*}
\end{center}
\caption{Reported average EPG for Pauli-randomized $\pi/2$ gates in different systems as determined by randomized benchmarking.}
\label{table:previous_benchmarking}
\end{table*}

To improve on the results of Ref.~\cite{knill08}, we integrated a
microwave antenna into a surface-electrode trap structure
\cite{ospelkaus08}. The use of microwave radiation instead of optical
stimulated-Raman transitions to drive qubit rotations suppresses
decoherence from laser beam pointing instability and power
fluctuations and eliminates decoherence from spontaneous emission. The
microwave amplitude can be stabilized more easily than laser power,
and because the antenna is integrated into the trap electrodes,
unwanted motion of the trap does not affect the microwave-ion-coupling
strength. The small distance (40~$\mu$m) between the trap surface and
the ion permits transition rates comparable to those based on
lasers. Improved shielding from ambient magnetic-field fluctuations
was achieved by locating the trap inside a copper vacuum enclosure
held at 4.2~K by a helium-bath cryostat. The thickness of the walls,
combined with the increase in electrical conductivity of copper at
4.2~K, effectively shields against the ambient magnetic field
fluctuations that typically limit coherence in room-temperature
ion-trap experiments \cite{knill08}. This shielding is evident when we
change the magnetic field external to the cryostat; the accompanying
response in ion fluorescence lags the change with an exponential time
constant of 3.8(2)~s. In addition, cryogenic operation decreases the
background gas pressure to negligible levels, thereby enabling long
experimental runs with the same ion, and it suppresses ion heating
\cite{deslauriers06,labaziewicz08,brown11}.

\begin{figure}
\centering \includegraphics{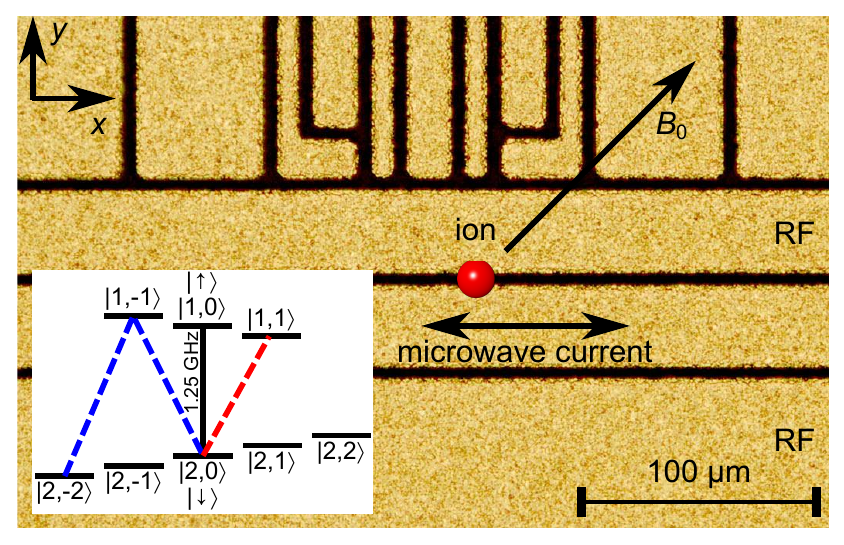}
\caption{(Color online) Micrograph of the ion trap, showing radio-frequency (rf)
    electrodes and control electrodes. The red sphere indicates the
    approximate ion position in the $x$-$y$ plane. Also shown are the
    directions of the static magnetic field $B_0$ and of the microwave
    current used to drive hyperfine transitions. (Inset)
    Energy level diagram (not to scale) of the 2s~$^2$S$_{1/2}$
    hyperfine states in $^9$Be$^+$. Blue dashed lines indicate
    the transitions used to prepare and measure
    $|\downarrow\rangle$. The solid black line indicates the qubit
    transition, and the red dashed line indicates one of the
    transitions used to shelve $|\uparrow\rangle$ into a dark state.}
\label{fig:schematic}
\end{figure}

The $^9$Be$^+$ ion is trapped $40$~$\mu$m above a surface-electrode
trap \cite{seidelin06} constructed of 8-$\mu$m-thick gold electrodes
electroplated onto a crystalline quartz substrate and separated by
5-$\mu$m gaps (Fig.~\ref{fig:schematic}).
A static magnetic field $B_0$, parallel to the trap surface and
collinear with a Doppler cooling laser beam, is applied to break the
degeneracy of the ground-state Zeeman sublevels
(Fig.~\ref{fig:schematic} inset). We drive 2s~$^2$S$_{1/2}$ hyperfine
transitions with microwave pulses near 1.25~GHz, coupled with a 4-nF
capacitor to one end of a trap control electrode. The microwave
current is shunted to ground at the other end of the electrode by the
4-nF capacitor of an RC filter. Microwave pulses are created by
frequency quadrupling the output of a direct-digital synthesizer whose
frequency and phase can be updated in less than 1~$\mu$s by a
field-programmable gate array [(FPGA), 16-ns timing resolution]. An rf
switch creates approximately rectangular-shaped pulses. This signal is
amplified and is delivered via a coaxial cable within the cryostat and
a feedthrough in the copper vacuum enclosure. In this Rapid
Communication, we use the clock transition
($|F=2,m_F=0\rangle\equiv|\downarrow\rangle
\leftrightarrow |1,0\rangle\equiv|\uparrow\rangle$) in $^9$Be$^+$ for
the qubit instead of the previously used $|2,-2\rangle \leftrightarrow
|1,-1\rangle$ transition \cite{knill08} (Fig.~\ref{fig:schematic}
inset). The clock transition is a factor of 20 less sensitive to
magnetic-field fluctuations (950~MHz/T at field
$B_0=1.51\times10^{-3}$~T, compared to 21~GHz/T).

A benchmarking experiment proceeds as follows. The ion is Doppler
cooled and optically pumped to the $|2,-2\rangle$ state with
$\sigma^{-}$-polarized laser radiation near the
$|^2$S$_{1/2},2,-2\rangle\leftrightarrow|^2$P$_{3/2},3,-3\rangle$
cycling transition at 313~nm. Then, the qubit is initialized in
$|\downarrow\rangle$ with two microwave $\pi$ pulses, resonant with
the $|2,-2\rangle\leftrightarrow|1,-1\rangle$ and
$|1,-1\rangle\leftrightarrow|\downarrow\rangle$ transitions (blue
lines in Fig.~\ref{fig:schematic} inset). Pulse duration is then
controlled by a digital delay generator, which has a 5-ps timing
resolution. The frequency, phase, and triggering of each pulse remain
under control of the FPGA.

A predetermined sequence of randomized computational gates is then
applied. Each computational gate consists of a Pauli gate
($\pi$ pulse) followed by a (non-Pauli) Clifford gate
($\pi/2$ pulse). The gate sequence is followed by measurement
randomization consisting of a random Pauli gate and a Clifford gate
chosen deterministically to yield an expected measurement outcome of
either $|\uparrow\rangle$ or $|\downarrow\rangle$. The Pauli gates are
chosen with equal probability from the set
$e^{-i\pi\sigma_\mathrm{p}/2}$, where $\sigma_\mathrm{p}
\in$~\{$\pm\sigma_x$, $\pm\sigma_y$, $\pm\sigma_z$, $\pm I$\}. The
Clifford gates are chosen with equal probability from the set
$e^{-i\pi\sigma_\mathrm{c}/4}$, where $\sigma_\mathrm{c}
\in$~\{$\pm\sigma_x$,~$\pm\sigma_y$\}. In practice, a Clifford gate is
implemented as a single (rectangular-shaped) $\pi/2$ pulse of duration
$\tau_{\pi/2}\approx 21$~$\mu$s with appropriate phase. For
calibration simplicity, a Pauli gate is implemented for
$\sigma_\mathrm{p}\in$ \{$\pm\sigma_x$, $\pm\sigma_y$\} as two
successive $\pi/2$ pulses, and an identity gate $\pm I$ is implemented
as an interval of the same duration without the application of
microwaves. A gate $e^{-i\pi\sigma_z/2}$ is implemented as an identity
gate, but the logical frame of the qubit and subsequent pulses are
adjusted to realize the relevant change in phase. All pulses are
separated by a delay of 0.72~$\mu$s.

\begin{figure*}
\centering \includegraphics{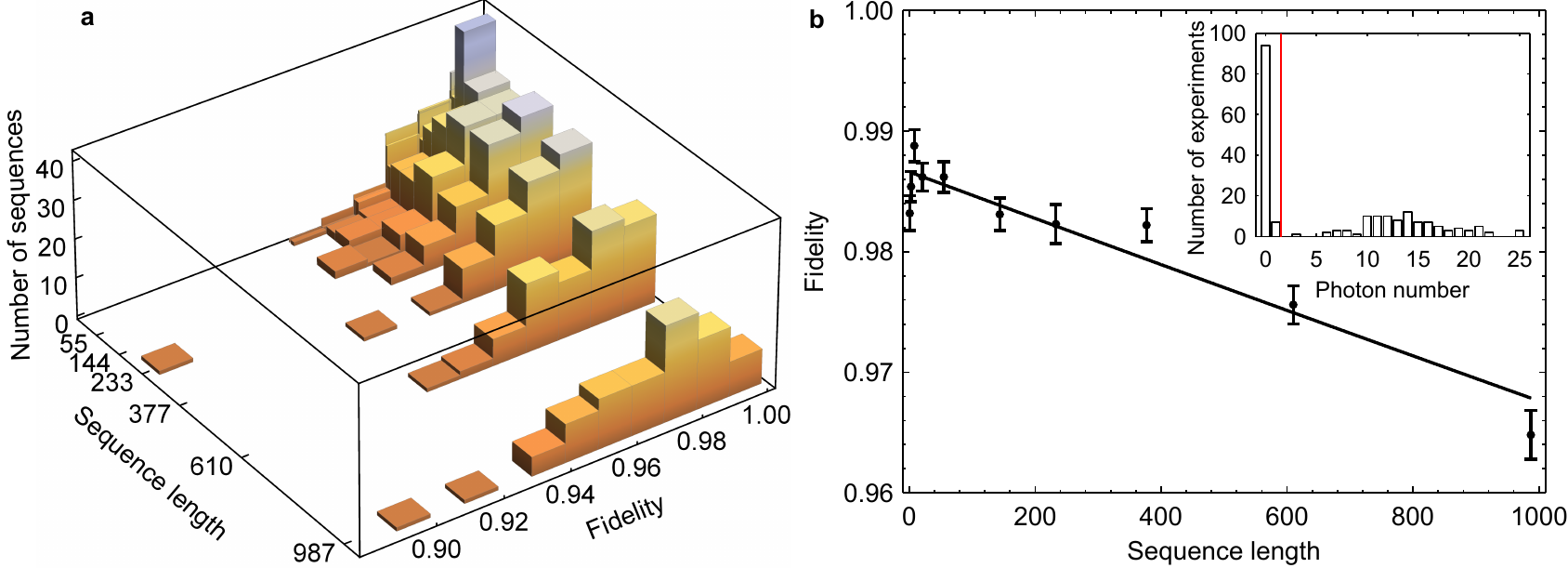}
\caption{(Color online) Results of the single-qubit benchmarking
    experiments. (a) Histogram of sequences of a given length
    with a given fidelity. Fidelity is discretized to 0.01 precision
    because 100 experiments were performed for each
    sequence. (b) Mean fidelity for each sequence length with
    error bars. The black trace is a least-squares fit to
    Eq.~(\ref{eqn:decay}) yielding an EPG of
    2.0(2)~$\times$~10$^{-5}$. (Inset) Summed histogram of
    bright and dark calibration experiments with a red line
    indicating the detection threshold.}
\label{fig:benchmark_result}
\end{figure*}

To detect the final qubit state, $\pi$ pulses implement the transfer
\mbox{$|\downarrow\rangle\rightarrow|1,-1\rangle\rightarrow|2,-2\rangle$}
(blue lines in Fig.~\ref{fig:schematic} inset). Two additional pulses
implement the transfer
\mbox{$|\uparrow\rangle\rightarrow|\downarrow\rangle\rightarrow|1,1\rangle$}
(black and red lines in Fig.~\ref{fig:schematic} inset). The ion is
then illuminated for 400~$\mu$s by 313-nm light resonant with the
cycling transition, and the resulting fluorescence is detected with a
photomultiplier. The entire sequence experiment (from initialization
through detection) is repeated 100 times (for each sequence) to reduce
statistical uncertainty. On average, approximately 13 photons are
collected from an ion in the bright $|2,-2\rangle$ state, but only
$0.14$ are collected from an ion in the dark $|1,1\rangle$ state
(due largely to laser light scattered from the trap surface). To
normalize the detection and to eliminate errors due to slow
fluctuations in laser power, each sequence experiment is immediately
followed by two reference experiments, where the ion is prepared in
the $|\downarrow\rangle$ and $|\uparrow\rangle$ states, respectively,
and the above detection protocol is implemented. From the resulting
bright and dark histograms [inset to
Fig.~\ref{fig:benchmark_result}(b)],
we take the median to establish a threshold for $|\downarrow\rangle$
and $|\uparrow\rangle$ detection.

Results are shown in Fig.~\ref{fig:benchmark_result}.
Sequence length refers to the number of computational gates in a
sequence. We implement sequences of lengths 1, 3, 8, 21, 55, 144, 233,
377, 610, and 987, with 100 different sequences at each length, for a
total of 1000 unique sequences. With the 21-$\mu$s $\pi/2$ duration
used here, a sequence of 987 computational gates requires
approximately 64~ms to complete. Our current software limits the
experiment to sequences of length $\lesssim1,300$~gates.

Theoretically, the average probability for obtaining a correct
measurement result (the fidelity) after a sequence of length $l$
is \cite{knill08}
\begin{equation}
\bar{\mathcal{F}}=\frac{1}{2}+\frac{1}{2}(1-d_\mathrm{if})(1-2 \mathcal{E}_\mathrm{g})^l,
\label{eqn:decay}
\end{equation}
where $d_\mathrm{if}$ describes errors in initialization and
measurement and $\mathcal{E}_\mathrm{g}$ is the EPG.  A
least-squares fit of the observed decay in fidelity to
Eq.~(\ref{eqn:decay}) yields
$\mathcal{E}_\mathrm{g}=2.0(2) \times 10^{-5}$ and
$d_\mathrm{if}=2.7(1) \times 10^{-2}$. Here, $d_\mathrm{if}$ is limited
by imperfect laser polarization caused by inhomogeneities in the
birefringence of the cryogenic windows of the vacuum enclosure.

The following systematic effects may contribute to the EPG:
magnetic-field fluctuations, microwave phase and frequency instability
and resolution limits, ac Zeeman shifts, pulse amplitude and
duration fluctuations, microwave-ion-coupling strength fluctuations,
decoherence caused by unintended laser illumination of the ion, and
off-resonant excitation to other levels in the ground-state hyperfine
manifold.

\begin{figure}
\centering \includegraphics{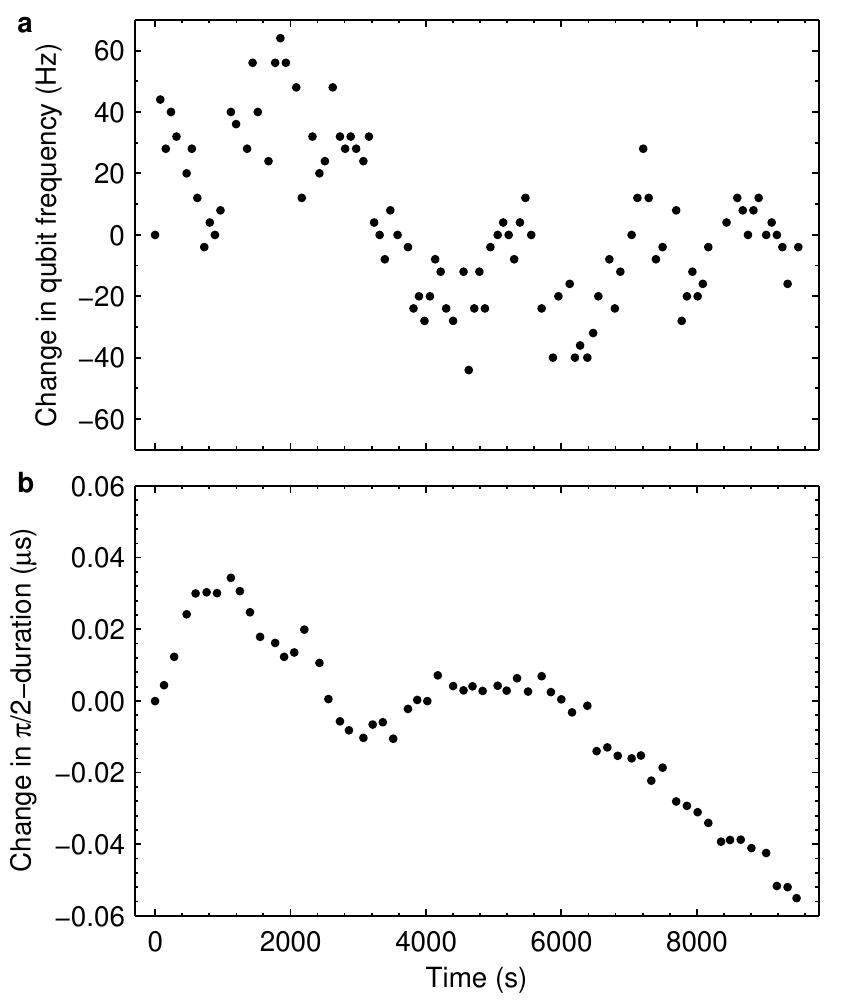}
\caption{Changes in (a) qubit transition frequency and (b)
  $\pi/2$ duration during the benchmarking experiments. Change is
  defined as the difference between the recalibrated value and the first
  calibration. Typical transition frequencies and $\pi/2$ durations are
  approximately 1.250~7385~GHz and 20.50~$\mu$s, respectively.}
\label{fig:calibration}
\end{figure}

During the benchmarking, we calibrate the qubit transition frequency
approximately every 60~s. The difference between each frequency
recalibration and the first calibration is plotted in
Fig.~\ref{fig:calibration}(a)
for the time period corresponding to the data in
Fig.~\ref{fig:benchmark_result}.
Monte Carlo simulations of the sequences indicate an EPG contribution
of $\mathcal{E}_\mathrm{g} =
\beta \Delta^2$, where $\beta=1.91\times10^{-8}/$Hz$^2$ and $\Delta$
is the detuning of the microwave frequency from the qubit frequency
(assumed constant for all of the sequences). In the absence of
recalibrations, the root-mean-square (rms) difference of 25~Hz would
give a predicted EPG of $1.2\times10^{-5}$. However, with regular
recalibration, the rms difference in frequency between adjacent
calibration points (15~Hz) gives a predicted contribution to the EPG
of $0.4\times10^{-5}$. The microwave frequency and phase
resolution are 0.37~Hz and $1.5$~mrad, respectively, leading to a
predicted EPG contribution of less than $10^{-7}$.

A theoretical estimate for the expected ac Zeeman shift of the clock
(qubit) transition yields a value of less than $1$~Hz. In principle,
this shift can be determined by comparing the qubit frequency measured
in a Ramsey experiment with that of a Rabi experiment. Such
back-to-back comparisons yielded values ranging from +14~Hz to -10~Hz,
each with errors of approximately 2~Hz. The source of this variation
and of the discrepancy with theory is not known, but if we assume, as
a worst-case scenario, a miscalibration of 15~Hz for the frequency, we
estimate an EPG contribution of $0.4\times10^{-5}$.

One measure of errors caused by qubit-frequency fluctuations
(e.g., from fluctuating magnetic fields) is to characterize decoherence
as an exponential decay through a T$_2$ process
\cite{olmschenk10,knill08}. To check this, we implement a Ramsey
experiment. The ion is prepared in $|\downarrow\rangle$, and we apply
a single $\pi/2$ pulse. After waiting for an interval $\tau/2$, we
apply a $\pi$ pulse to refocus the qubit, and following another
interval $\tau/2$, we apply an additional $\pi/2$ pulse, ideally
restoring the qubit back to $|\downarrow\rangle$. An exponential fit
of the resulting decay in the $|\downarrow\rangle$ state probability over
periods $\tau\lesssim100$~ms gives T$_2=0.38(4)$~s. Assuming this
value of T$_2$ also describes frequency fluctuations at times on the
order of the gate pulses, we predict an EPG contribution of
$9\times10^{-5}$. Because this exceeds the benchmark value, we believe
that the noise at shorter periods, in this experiment, is smaller than
that predicted by a simple exponential fitted at longer durations.

We recalibrate the $\pi/2$ duration approximately every 120~s with a
sequence of 256 in-phase $\pi/2$ pulses [Fig.~\ref{fig:calibration}(b)].
Monte Carlo simulations indicate an EPG contribution of
$\mathcal{E}_\mathrm{g} = \gamma (\Delta\tau)^2$, where
$\gamma=2.7\times10^{-3}/\mu$s$^2$ and $\Delta\tau$ represents a
miscalibration in the $\pi/2$ time (assumed constant for all
sequences).  In the absence of recalibration, the 23-ns rms drift
would correspond to an EPG of $0.1\times10^{-5}$; from the estimated
residual miscalibration between points of 5 ns, we predict an EPG
contribution of less than $10^{-7}$.

We characterize pulse-to-pulse microwave power fluctuations by turning
on the microwaves continuously and sampling the power every 10~ns. The
integral of the sampled power over a 25-$\mu$s interval is
proportional to the total rotation angle during a pulse of the same
duration. We perform this integral 12 times, with each 25-$\mu$s
interval following the previous one by 10~s. Within 120~s after turning on
the microwaves, we observe a 1\% drift in the power. If the
pulse-to-pulse variation in microwave power is, in fact, this large, it
corresponds to an EPG contribution of $3\times10^{-5}$. However, after
a 20-min warm-up interval, we measure a pulse-to-pulse power
variation of only 0.1\%, corresponding to an EPG contribution of
$0.03\times10^{-5}$. Because the duty cycle of the benchmarking
experiment is not constant, with sequences of different lengths and
calibration experiments interspersed throughout, it is difficult to
assign a specific EPG contribution to this effect. However, we do
observe larger EPG at higher microwave powers, consistent with
temperature effects playing a role at these higher powers.

To investigate unintended laser light as a source of decoherence,
(e.g., from optical pumping), the ion is prepared in
$|\downarrow\rangle$ and is allowed to remain in the dark for
varying durations. We observe no decay in the $|\downarrow\rangle$
state probability with an uncertainty of $2\times10^{-7}$/$\mu$s,
corresponding to the absence of gate errors during the 65-$\mu$s
randomized gate interval with an uncertainty of
$1\times10^{-5}$. Similar results are obtained for an ion prepared in
$|\uparrow\rangle$.

Microwave-induced transitions from the qubit levels into other Zeeman
levels within the ground-state hyperfine manifold can be inferred by
observing an asymmetry between sequences ending in
$|\downarrow\rangle$ and those ending in $|\uparrow\rangle$. While the
$|2,-2\rangle$ state fluoresces with 13 photons detected on average,
other hyperfine states yield, at most, 1.3 photons during the 400-$\mu$s
detection period. Therefore, transitions from the qubit manifold to
other levels would show up as a loss of fidelity for sequences ending
in $|\downarrow\rangle$, while they would not affect the apparent
fidelity of sequences ending in $|\uparrow\rangle$. For the bright
sequences in Fig.~\ref{fig:benchmark_result}, the EPG is
$2.2(5)\times10^{-5}$, while for the dark sequences it is
$2.0(5)\times10^{-5}$. We conclude that qubit leakage contributes an
EPG of $<0.2(7)\times10^{-5}$. Similarly, if ion heating contributes
to the EPG, it should appear as a deviation from exponential decay in
the benchmarking data, which we do not observe.

For future work, it seems likely that microwave power fluctuations
could be controlled passively through a suitable choice of amplifiers
and switching circuitry or actively via feedback. Shorter pulses at
higher microwave powers would diminish errors associated with
fluctuating qubit frequency, but errors due to off-resonant
transitions become more of a concern in this regime. Off-resonant
transitions could be suppressed with the use of appropriately shaped
pulses, which concentrate the microwave spectrum near the qubit
transition frequency. Self-correcting pulse sequences
\cite{levitt86} could be used to reduce the effects of errors in
$\pi/2$ duration and transition frequency. In a multizone trap array,
single-qubit gates implemented with microwaves will be susceptible to
cross talk between zones; however, this effect can be mitigated with
careful microwave design, the use of nulling currents in spectator
zones \cite{ospelkaus08}, and the use of composite pulses
\cite{levitt86}. A demonstration of two-qubit gates with errors small
enough to enable scalable quantum computing remains challenging, but
high-fidelity single-qubit gates should make this task easier. For
example, many refocusing and decoupling techniques are based on
single-qubit gates and can reduce errors during two-qubit gate
operations \cite{viola99}.

\begin{acknowledgments}
This work was supported by IARPA, NSA, DARPA, ONR, and the NIST
Quantum Information Program. We thank U. Warring, M. Biercuk,
A. VanDevender, J. Amini, and R. B. Blakestad for their help in
assembling parts of the experiment, and we thank J. Britton,
S. Glancy, A. Steane, and C. Bennett for comments. This article is a
contribution of the U.S. Government, not subject to U.S. copyright.
\end{acknowledgments}


\begin{thebibliography}{22}
\expandafter\ifx\csname natexlab\endcsname\relax\def\natexlab#1{#1}\fi
\expandafter\ifx\csname bibnamefont\endcsname\relax
  \def\bibnamefont#1{#1}\fi
\expandafter\ifx\csname bibfnamefont\endcsname\relax
  \def\bibfnamefont#1{#1}\fi
\expandafter\ifx\csname citenamefont\endcsname\relax
  \def\citenamefont#1{#1}\fi
\expandafter\ifx\csname url\endcsname\relax
  \def\url#1{\texttt{#1}}\fi
\expandafter\ifx\csname urlprefix\endcsname\relax\def\urlprefix{URL }\fi
\providecommand{\bibinfo}[2]{#2}
\providecommand{\eprint}[2][]{\url{#2}}

\bibitem[{\citenamefont{Nielsen and Chuang}(2000)}]{nielsen01}
\bibinfo{author}{\bibfnamefont{M.~A.} \bibnamefont{Nielsen}} \bibnamefont{and}
  \bibinfo{author}{\bibfnamefont{I.~L.} \bibnamefont{Chuang}},
  \emph{\bibinfo{title}{Quantum Computation and Quantum Information}}
  (\bibinfo{publisher}{Cambridge University Press},
  \bibinfo{address}{Cambridge, UK}, \bibinfo{year}{2000}).

\bibitem[{\citenamefont{Ladd et~al.}(2010)}]{ladd10}
\bibinfo{author}{\bibfnamefont{T.~D.} \bibnamefont{Ladd}} \bibnamefont{et~al.},
  \bibinfo{journal}{Nature (London)} \textbf{\bibinfo{volume}{464}}, \bibinfo{pages}{45}
  (\bibinfo{year}{2010}).

\bibitem[{\citenamefont{DiVincenzo}(2000)}]{divincenzo00}
\bibinfo{author}{\bibfnamefont{D.~P.} \bibnamefont{DiVincenzo}},
  \bibinfo{journal}{Fortschr. Phys.} \textbf{\bibinfo{volume}{48}},
  \bibinfo{pages}{771} (\bibinfo{year}{2000}).

\bibitem[{\citenamefont{Preskill}(1998)}]{preskill98}
\bibinfo{author}{\bibfnamefont{J.}~\bibnamefont{Preskill}},
  \bibinfo{journal}{Proc. R. Soc. Lond., Ser. A} \textbf{\bibinfo{volume}{454}},
  \bibinfo{pages}{385} (\bibinfo{year}{1998}).

\bibitem[{\citenamefont{Knill}(2010)}]{knill10}
\bibinfo{author}{\bibfnamefont{E.}~\bibnamefont{Knill}},
  \bibinfo{journal}{Nature (London)} \textbf{\bibinfo{volume}{463}},
  \bibinfo{pages}{441} (\bibinfo{year}{2010}).

\bibitem[{\citenamefont{Ryan et~al.}(2009)\citenamefont{Ryan, Laforest, and
  Laflamme}}]{ryan09}
\bibinfo{author}{\bibfnamefont{C.~A.} \bibnamefont{Ryan}},
  \bibinfo{author}{\bibfnamefont{M.}~\bibnamefont{Laforest}}, \bibnamefont{and}
  \bibinfo{author}{\bibfnamefont{R.}~\bibnamefont{Laflamme}},
  \bibinfo{journal}{New J. Phys.} \textbf{\bibinfo{volume}{11}},
  \bibinfo{pages}{013034} (\bibinfo{year}{2009}).

\bibitem[{\citenamefont{Olmschenk et~al.}(2010)}]{olmschenk10}
\bibinfo{author}{\bibfnamefont{S.}~\bibnamefont{Olmschenk}}
  \bibnamefont{et~al.}, \bibinfo{journal}{New J. Phys.}
  \textbf{\bibinfo{volume}{12}}, \bibinfo{pages}{113007}
  (\bibinfo{year}{2010}).

\bibitem[{\citenamefont{Benhelm et~al.}(2008)}]{benhelm08}
\bibinfo{author}{\bibfnamefont{J.}~\bibnamefont{Benhelm}} \bibnamefont{et~al.},
  \bibinfo{journal}{Nat. Phys.} \textbf{\bibinfo{volume}{4}},
  \bibinfo{pages}{463} (\bibinfo{year}{2008}).

\bibitem[{\citenamefont{Ospelkaus et~al.}(2011)}]{ospelkaus11}
\bibinfo{author}{\bibfnamefont{C.}~\bibnamefont{Ospelkaus}}
  \bibnamefont{et~al.}, \bibinfo{journal}{Nature (London)} \textbf{\bibinfo{volume}{476}},
  \bibinfo{pages}{181} (\bibinfo{year}{2011}).

\bibitem[{\citenamefont{Knill et~al.}(2008)}]{knill08}
\bibinfo{author}{\bibfnamefont{E.}~\bibnamefont{Knill}} \bibnamefont{et~al.},
  \bibinfo{journal}{Phys. Rev. A} \textbf{\bibinfo{volume}{77}},
  \bibinfo{pages}{012307} (\bibinfo{year}{2008}).

\bibitem[{\citenamefont{Poyatos et~al.}(1997)\citenamefont{Poyatos, Cirac, and
  Zoller}}]{poyatos97}
\bibinfo{author}{\bibfnamefont{J.~F.} \bibnamefont{Poyatos}},
  \bibinfo{author}{\bibfnamefont{J.~I.} \bibnamefont{Cirac}}, \bibnamefont{and}
  \bibinfo{author}{\bibfnamefont{P.}~\bibnamefont{Zoller}},
  \bibinfo{journal}{Phys. Rev. Lett.} \textbf{\bibinfo{volume}{78}},
  \bibinfo{pages}{390} (\bibinfo{year}{1997}).

\bibitem[{\citenamefont{Bravyi and Kitaev}(2005)}]{bravyi05}
\bibinfo{author}{\bibfnamefont{S.}~\bibnamefont{Bravyi}} \bibnamefont{and}
  \bibinfo{author}{\bibfnamefont{A.}~\bibnamefont{Kitaev}},
  \bibinfo{journal}{Phys. Rev. A} \textbf{\bibinfo{volume}{71}},
  \bibinfo{pages}{022316} (\bibinfo{year}{2005}).

\bibitem[{\citenamefont{Knill}(2005)}]{knill05}
\bibinfo{author}{\bibfnamefont{E.}~\bibnamefont{Knill}},
  \bibinfo{journal}{Nature (London)} \textbf{\bibinfo{volume}{434}}, \bibinfo{pages}{39}
  (\bibinfo{year}{2005}).

\bibitem[{\citenamefont{Biercuk et~al.}(2009)}]{biercuk09}
\bibinfo{author}{\bibfnamefont{M.~J.} \bibnamefont{Biercuk}}
  \bibnamefont{et~al.}, \bibinfo{journal}{Quantum Inf. Comput.}
  \textbf{\bibinfo{volume}{9}}, \bibinfo{pages}{920} (\bibinfo{year}{2009}).

\bibitem[{\citenamefont{Chow et~al.}(2010)}]{chow10}
\bibinfo{author}{\bibfnamefont{J.~M.} \bibnamefont{Chow}} \bibnamefont{et~al.},
  \bibinfo{journal}{Phys. Rev. A} \textbf{\bibinfo{volume}{82}},
  \bibinfo{pages}{040305} (\bibinfo{year}{2010}).

\bibitem[{\citenamefont{Ospelkaus et~al.}(2008)}]{ospelkaus08}
\bibinfo{author}{\bibfnamefont{C.}~\bibnamefont{Ospelkaus}}
  \bibnamefont{et~al.}, \bibinfo{journal}{Phys. Rev. Lett.}
  \textbf{\bibinfo{volume}{101}}, \bibinfo{pages}{090502}
  (\bibinfo{year}{2008}).

\bibitem[{\citenamefont{Deslauriers et~al.}(2006)}]{deslauriers06}
\bibinfo{author}{\bibfnamefont{L.}~\bibnamefont{Deslauriers}}
  \bibnamefont{et~al.}, \bibinfo{journal}{Phys. Rev. Lett.}
  \textbf{\bibinfo{volume}{97}}, \bibinfo{pages}{103007}
  (\bibinfo{year}{2006}).

\bibitem[{\citenamefont{Labaziewicz et~al.}(2008)}]{labaziewicz08}
\bibinfo{author}{\bibfnamefont{J.}~\bibnamefont{Labaziewicz}}
  \bibnamefont{et~al.}, \bibinfo{journal}{Phys. Rev. Lett.}
  \textbf{\bibinfo{volume}{100}}, \bibinfo{pages}{013001}
  (\bibinfo{year}{2008}).

\bibitem[{\citenamefont{Brown et~al.}(2011)}]{brown11}
\bibinfo{author}{\bibfnamefont{K.~R.} \bibnamefont{Brown}}
  \bibnamefont{et~al.}, \bibinfo{journal}{Nature (London)}
  \textbf{\bibinfo{volume}{471}}, \bibinfo{pages}{196} (\bibinfo{year}{2011}).

\bibitem[{\citenamefont{Seidelin et~al.}(2006)}]{seidelin06}
\bibinfo{author}{\bibfnamefont{S.}~\bibnamefont{Seidelin}}
  \bibnamefont{et~al.}, \bibinfo{journal}{Phys. Rev. Lett.}
  \textbf{\bibinfo{volume}{96}}, \bibinfo{pages}{253003}
  (\bibinfo{year}{2006}).

\bibitem[{\citenamefont{Levitt}(1986)}]{levitt86}
\bibinfo{author}{\bibfnamefont{M.~H.} \bibnamefont{Levitt}},
  \bibinfo{journal}{Prog. Nucl. Magn. Reson. Spectrosc.}
  \textbf{\bibinfo{volume}{18}}, \bibinfo{pages}{61} (\bibinfo{year}{1986}).

\bibitem[{\citenamefont{Viola et~al.}(1999)\citenamefont{Viola, Lloyd, and
  Knill}}]{viola99}
\bibinfo{author}{\bibfnamefont{L.}~\bibnamefont{Viola}},
  \bibinfo{author}{\bibfnamefont{S.}~\bibnamefont{Lloyd}}, \bibnamefont{and}
  \bibinfo{author}{\bibfnamefont{E.}~\bibnamefont{Knill}},
  \bibinfo{journal}{Phys. Rev. Lett.} \textbf{\bibinfo{volume}{83}},
  \bibinfo{pages}{4888} (\bibinfo{year}{1999}).

\end{thebibliography}
\end{document}